\def\year{2024}\relax
\documentclass[letterpaper]{article} % DO NOT CHANGE THIS
\usepackage{aaai22}  % DO NOT CHANGE THIS
\usepackage{times}  % DO NOT CHANGE THIS
\usepackage{helvet}  % DO NOT CHANGE THIS
\usepackage{courier}  % DO NOT CHANGE THIS
\usepackage[hyphens]{url}  % DO NOT CHANGE THIS
\usepackage{graphicx} % DO NOT CHANGE THIS
\usepackage{natbib}  % DO NOT CHANGE THIS AND DO NOT ADD ANY OPTIONS TO IT
\usepackage{caption} % DO NOT CHANGE THIS AND DO NOT ADD ANY OPTIONS TO IT
\DeclareCaptionStyle{ruled}{labelfont=normalfont,labelsep=colon,strut=off} % DO NOT CHANGE THIS
\usepackage{algorithm,setspace}
\usepackage[noend]{algorithmic}

\usepackage{dblfloatfix}
\usepackage{placeins}

\usepackage{amsmath}
\usepackage{booktabs}

\usepackage{xcolor} % TOREMOVE 

\usepackage{newfloat}
\usepackage{listings}
\floatstyle{ruled}
\newfloat{listing}{tb}{lst}{}
\floatname{listing}{Listing}
\title{Unmasking information manipulation: A quantitative approach to detecting Copy-pasta, Rewording, and Translation on Social Media}
\author {
    Manon Richard \footnote{These authors contributed equally to the work},
    Lisa Giordani \textsuperscript{*},
    Cristian Brokate,
    Jean Liénard \footnote{Correspondence should be addressed to \texttt{jean.lienard@sgdsn.gouv.fr}}
}
\affiliations {
    VIGINUM, Secrétariat Général de la Sécurité et de la Défense Nationale, France
}

\usepackage{bibentry}
\begin{document}

\maketitle

\begin{abstract}

This study proposes a comprehensive methodology for identifying three techniques utilized in foreign-operated information manipulation campaigns: Copy-Pasta, Rewording, and Translation. Our approach, dubbed the ``$3\Delta$-space duplicate methodology'', quantifies the semantic, grapheme, and language aspects of messages. Computing pairwise distances within these dimensions enables detection of abnormally close messages that are likely part of a coordinated campaign. We validate our approach using a synthetic dataset generated with ChatGPT and DeepL, further applying it to a real-world dataset on Venezuelan actors from Twitter Transparency. Our method successfully identifies all three types of inauthentic duplicates in the synthetic dataset, and is able to uncover inauthentic duplicates across political, commercial, and entertainment contexts in the Twitter dataset. The distinct focus on clustered alterations to messages, rather than individual messages, makes our approach efficient and effective at detecting large-scale instances of textual manipulation, including AI-generated ones. Moreover, our method offers a robust tool for identifying translated content, overlooked in previous research. This research also represents the first comprehensive analysis of copy-pasta detection, providing a reliable technique for tracking duplicate textual content across social networks.

\end{abstract}

\noindent The primary aim of an information manipulation campaign is to influence the opinions of individuals through various strategies and tactics, with a dedicated team orchestrating the campaign and an underlying intent of manipulation \cite{bradshaw2018challenging}. This endeavor involves disseminating a set of facts or opinions to a specific target audience, encompassing distinct phases, including identifying the relevant audience, studying their vulnerabilities to information, crafting the core message, executing the technical steps for message propagation, and evaluating the campaign’s outcome.

Of particular interest is the examination of how a specific language element, such as a concise and impactful narrative, can be transformed to effectively reach its target audience. While duplicating content verbatim can be efficient, it faces two fundamental limitations: linguistic barriers that confine the targeted audience to a limited geographic scope, and platform moderation, capable of both flagging and regulating message dissemination. It also faces the risk of being easily identified and exposed publicly by actors of the civil society. Thus, the imperative arises for the adoption of advanced techniques.

While extensive research has focused on artificial amplification, there has been limited exploration into how textual content is manipulated to circumvent platform moderation. \emph{Copy-pasta}, derived from ``copy-paste'' \cite{turton2015trufax}, refers to a range of techniques involving duplicating content, often with minor modifications like adding hashtags, emojis, or altering a single word. This strategy has received scant attention in academic literature, with only a few notable exceptions like \citet{{suresh2023tracking}}. Empirical investigations, such as the one conducted by \citet{DFRLab2019}, often rely on makeshift methods to identify this type of content duplication.

While copy-pasta stands out as the quickest and simplest procedure, \emph{Rewording} is a more sophisticated technique that traditionally required language-fluent agents. However, the advent of AI-generated chatbots, such as ChatGPT, has made it prevalent in social networks \cite{ferrara2023social}. Yet, existing methods for distinguishing AI-generated text from innocuous content remain inadequate \cite{yang2023anatomy}.

\emph{Translation} demands access to reliable translation tools or bilingual agents to credibly adapt a collection of messages for a different language audience. Although it can be relatively costly to execute convincingly, it holds particular significance in the context of foreign-operated campaigns. In such campaigns, narratives may be crafted in the perpetrator's native language or a widely understood language like English, while targeting a diverse audience, including e.g. speakers of various European languages. Despite its relevance, there is currently no established method for detecting translations within a dataset.

In this study, we introduce a comprehensive approach for identifying three primary methods -- Copy-Pasta, Rewording, and Translation -- commonly employed in foreign-operated information manipulation campaigns to alter language elements. Our approach centers on labeling message pairs based on the applied method, utilizing three distinct dimensions: message meaning (semantic), specific wording represented by graphemes, and the language in which the message is written. We calculate pairwise distances within each of these dimensions ($\Delta_{\mbox{semantic}}$, $\Delta_{\mbox{grapheme}}$ and $\Delta_{\mbox{language}}$) to identify abnormally close messages. This methodology, which we term the ``$3\Delta$-space duplicate methodology'', forms the core of this research.

We validate the robusteness of our approach through two distinct datasets. First, we crafted a synthetic dataset that encompasses the transformations in the $3\Delta$-space by means of a generative language AI (ChatGPT) as well as an automatic translation tool (DeepL). We use this dataset to benchmark several candidate algorithms to compute $\Delta_{\mbox{semantic}}$ and $\Delta_{\mbox{grapheme}}$. Second, we apply our integrated approach to a real-world Twitter dataset related to Venezuelan actors, released by Twitter Transparency in 2021. Our methodology successfully identifies all three types of inauthentic duplicates: those boosting political messages, commercial messages for alcoholic beverages, and messages related to the TV show ``The Walking Dead,'' possibly intended to conceal other activities or target specific audiences. Furthermore, our analysis reveals two distinct account typologies employing different strategies. One group exclusively employs Copy-Pasta and Rewording techniques, while the other relies solely on Rewording and Translation methods. This insight could shed light on the presence of different teams or groups involved in these types of campaigns.

%%%%%%%%%%%%%%%%
%% DEFINITION %%
%%%%%%%%%%%%%%%%
\section{Related work}

Coordinated Inauthentic Behavior (CIB) refers to the synchronized efforts of individuals or groups to shape online discourse through repeated and synchronized activity. As \citet{weber2021amplifying} puts it, identifying CIB is akin ``to identify groups of accounts whose behaviour, though typical in nature, is anomalous in degree.'' Early computational work in CIB detection has predominantly relied on supervised detection frameworks \cite{cresci2020decade}, where algorithms are trained to detect specific account or message characteristics, such as globally stereotyped posting activity in account timelines \cite{cresci2016dna} or aggregating numerous features of a single account to establish its artificial degree \cite{davis2016botornot}.

More recently, unsupervised graph-based approaches have achieved superior performances in identifying CIB \cite{pacheco2021uncovering,nizzoli2021coordinated,weber2021amplifying}. Formally, these approaches involve computing a distance between messages and cluster them right away \cite{pacheco2021uncovering} or after additional processing \cite{nizzoli2021coordinated}. In each of these instances, the outcome is a cluster of messages or accounts that are synchronized based on a particular dimension. This cluster can then be subjected to analysis, aiding in the comprehension of the content of the information manipulation campaign.

Notably, \citeauthor{suresh2023tracking} (\citeyear{suresh2023tracking}) and \citeauthor{ng2022cross} (\citeyear{ng2022cross}) have developed frameworks for detecting duplicated content in a graph-based approach. \citet{suresh2023tracking} proposed using the Ratclif-Obershelp distance to detect copy-pasta, similar to the Grapheme-distance we employ in conjunction with the Language-distance. \citet{ng2022cross} developed a sophisticated methodology to identify repeated topics in YouTube video transcripts based on word-level embeddings, which are then consolidated at the video level and assessed for pairwise distance. Our approach however directly utilizes sentence-level embeddings, which simplifies the computation of the Semantic-distance step considerably.

Besides research focusing on CIB, there exists a substantial body of work dedicated to plagiarism detection, where an active research community focuses on methods for identifying duplicated documents or sentences \cite{jiffriya2021plagiarism}. In fact, the individual components we employ have been subjects of separate studies in the context of plagiarism detection, including language-invariant embeddings \cite{alotaibi2020using} as well as string-based and semantic-based proximity \cite{sunilkumar2019survey,khaled2021plagiarism}. Our research can be seen as a specialized extension of this broader field. We specifically target very short texts, which are typical of social media content. Additionally, our interest goes beyond simply determining if a document is duplicated; we aim to understand the methods employed for duplication. Furthermore, we focus on network-level analyses, where our ultimate goal is to uncover and analyze clusters of duplicated messages and the responsible accounts.

The specific dataset employed to validate our methodology has undergone prior examination. \citet{stanfordvenez} use network analyses based on user mentions and shared hashtags. Their investigation resulted in the identification of four to five distinct user clusters characterized by coordinated behaviors, indicative of potential involvement in commercial tweet-for-hire arrangements. In contrast, our approach detects the user clusters and offers a finer level of detail regarding the techniques employed for disseminating coordinated campaigns. 

%%%%%%%%%%%%%%%%%
%% METHODOLOGY %%
%%%%%%%%%%%%%%%%%
\section{Methodology}

\subsection{Overview of the approach}

The $3\Delta$-space duplicate methodology identifies translations, rewordings, and copy-pasta, through an iterative decision process involving the following steps:

\begin{enumerate}
    \item quantifying the \textit{semantic proximity} of messages, which encapsulates the meaning of messages
    \item for messages with close semantic proximity, we further identify their \textit{language proximity}, and label as ``translation'' those expressing the same meaning in different languages 
    \item for messages with close semantic proximity and the same language, we finally quantify their \textit{grapheme proximity} as the number of letter changes to move from one to another: minor changes are labelled as ``copy-pasta'', and large changes as ``rewording''
\end{enumerate}

The general outline of this decision process is formalized in Algorithm \ref{alg:overview}. Our algorithm thus relies on computing three different distances, which we note $\Delta_{\mbox{semantic}}$, $\Delta_{\mbox{grapheme}}$ and $\Delta_{\mbox{language}}$. Those distances are designed to fall within the $[0, 1]$ range, where low values depict close messages. The overall process involves to compute those distances on all pairs ($X_1$, $X_2$) of messages in the dataset.

\begin{algorithm}[tb]
\caption{Global inference algorithm}
\label{alg:overview}
\begin{algorithmic}[1] %[1] enables line numbers
\setstretch{1.35}
\STATE Let $X_1, X_2$ be all pairs from a corpus. 
\IF{$\Delta_{\mbox{grapheme}}(X_1, X_2) < \tau_p$}
    \STATE \textbf{return ``Copy-Pasta''}
\ELSIF{$\Delta_{\mbox{semantic}}(X_1, X_2) < \tau_s$}
        \IF{$\Delta_{\mbox{language}}(X_1, X_2) < \tau_l$}
            \STATE \textbf{return ``Rewording''}
        \ELSE
           \STATE \textbf{return ``Translation''}
        \ENDIF
\ELSE
    \STATE \textbf{return ``No match''}
\ENDIF
\end{algorithmic}
\end{algorithm}

\subsection{Semantic proximity}

We employ two popular sentence-level embeddings in our quest to develop a method for extracting the semantic essence from short texts, irrespective of their language or wording.

We first test the Universal Sentence Encoder (USE) \cite{yang2019multilingual} to compute an embedding for each document, in the form of a vector of 512 values. This encoder was trained on a large corpus mined from various online sources including Reddit, StackOverflow, and YahooAnswers, as well as the Stanford Natural Language Inference dataset \cite{bowman2015large}. This dataset features 16 widely used languages, with at least 60 million question-pairs per language, resulting in more than a billion question-answers pairs overall \cite{yang2019multilingual}. Although it was originally developed for single sentences, it can also be applied on short documents containing two to three sentences, thus making it amenable to work with tweets and other short texts. By the choice of the training tasks, the USE embedding is trained to convey the semantic meaning of the sentence, independently of its precise wording or language. It should thus be invariant to rewording, copy-pasta and translations. This model is freely available on Tensorflow Hub.\footnote{\url{https://tfhub.dev/google/universal-sentence-encoder-multilingual-large/3}}

As an alternative, we evaluate OpenAI's \texttt{text-embedding-ada-002} embedding provider. Details on the implementation and training data of this model are sparse, but it is reported to work in cross-language settings and performs well on a varieties of benchmarks\footnote{\url{https://openai.com/blog/new-and-improved-embedding-model}}. It is currently available for a fee using OpenAI's API. 

Regardless of the embedding chosen, we compute distances for any two given embeddings, $E_1$ and $E_2$ of strings $X_1$ and $X_2$, using the same formula as in \citet{cer2018universal}: 

\begin{equation}
\Delta_{\mbox{semantic}}(X_1, X_2) = \frac{1}{\pi}\ arccos(\frac{E_1\ .\ E_2}{||E_1||\ ||E_2||})
\label{eq:dist}
\end{equation}

This is in fact an angular distance ($\frac{1}{\pi}\ arccos$) derived from the cosine similarity ($\frac{X_1\ .\ X_2}{||X_1||\ ||X_2||}$), which has the main advantage of yielding normalized values in the $[0, 1]$ range, with low values corresponding to closely related message meanings. Alternative distances such as $L2$-norm were also tried with very similar performance outcomes.

\subsection{Language proximity}

Differentiating rewording from translation involves inferring the language of a given document. Here we compute language distance as a straightforward binary measure:

\begin{equation}
  \Delta_{\mbox{language}}(X_1, X_2) =  
    \begin{cases}
      0, & \text{if}\ lang(X_1) = lang(X_2) \\
      1, & \text{otherwise}
    \end{cases}
\end{equation}

The language identifier  $lang()$ can be obtained from various sources. In this work, we utilize the platform-provided data from the Twitter Transparency dataset and resort to the original dataset label in synthetic experiments. In a broader context, one can calculate it using specialized tools such as Fasttext \cite{joulin2016fasttext}, which is particularly well-suited as it was trained on short messages commonly found in social media interactions and supports a wide range of languages \cite{haider2023detecting}.

\subsection{Grapheme proximity}

In this work, our definition of Copy-Pasta encompasses both exact duplication as well as minor textual alterations, such as the inclusion of emojis, punctuation modifications, spacing adjustments, single-word substitutions, the addition of introductory headers like ``Breaking News!'', or the appending of sets of hashtags at the end of a message. Messages sharing the same language and meaning, yet consisting of highly similar strings, are categorized as ``Copy-Pasta'', while those displaying substantial differences in their constituent letters or graphemes fall into the ``Rewording'' category.

We benchmark three algorithms measuring grapheme distances, including the classical distances of Levenshtein \cite{1966SPhD...10..707L} and Ratclif-Obershelp \cite{ratclif}, as well as the newer $gzip$-compressor distance \cite{jiang-etal-2023-low}. We also test two original grapheme proximity measures based on the distance between bag-of-bigrams vectors.

To account for different message length in our dataset, we normalize Levenshtein distance between two strings, $lv(X_1, X_2)$, based on the longest string:

\begin{equation}
  \Delta_{\mbox{grapheme}}^{lv}(X_1, X_2) = 1 - \frac{lv(X_1, X_2)}{max(len(X_1),\ len(X_2))}
\end{equation}

The maximal Levenshtein distance between two strings is equal to the length of the longest string, which implies that it yields values in the $[0, 1]$ range.

Likewise, we normalize the Ratclif-Obershelp distance $ro(X_1, X_2)$ to also take values in $[0, 1]$:

\begin{equation}
  \Delta_{\mbox{grapheme}}^{ro}(X_1, X_2) = 2\ \frac{ro(X_1, X_2)}{len(X_1)+len(X_2)}
\end{equation}

Following \citet{jiang-etal-2023-low}, the normalized version of the $gzip$-compressor distance is computed as:

\begin{multline}
  \Delta_{\mbox{grapheme}}^{gz}(X_1, X_2) =  \\
1 - \frac{gz(X_1\ X_2) - min(gz(X_1), gz(X_2))}{max(gz(X_1), gz(X_2))}
\end{multline}

Where $gz()$ refers to the $gzip$ compression, and $X_1\ X_2$ is the string resulting from the concatenation of $X_1$ and $X_2$.

We also propose two measures of string proximity based on bigrams. The first measure is based on bigrams of words (i.e., ``$two$ $words$'' is considered as the bigram (``$two$'', ``$words$'')), and the second on bigrams of characters (``$ab$'' is then counted as  (``$a$'', ``$b$''). In both instances, we tabulate the occurrences of all bigrams present in at least one string. This means to identify the $n$ unique bigrams (``$a$'', ``$b$'') present in at least one of the two strings $X_1$ and $X_2$, and then tally their occurrences in the strings:

\begin{equation}
    bg_i(``a", ``b") =
    \begin{cases}
      0 \text{ if ``ab'' is not present in } X_i \\
      \text{number of times ``ab'' is present otherwise}
    \end{cases}
\end{equation}

We further compute $\Delta_{\mbox{grapheme}}^{bg}(X_1, X_2)$ as the absolute scaled distance between these vectors:

\begin{multline}
\Delta_{\mbox{grapheme}}^{bg}(X_1, X_2) = \\
\frac{\sum_{(``a", ``b") \in X_1 \cup X_2}{|bg_1(``a", ``b") - bg_2(``a", ``b")}|}{\sum_{(``a", ``b") \in X_1 \cup X_2}{bg_1(``a'', ``b") + bg_2(``a", ``b")}}
\end{multline}

In the model comparison, we further label $\Delta_{\mbox{grapheme}}^{bg-w}$ the distance computed on word bigrams, and $\Delta_{\mbox{grapheme}}^{bg-l}$ the distance computed on letter bigrams.

\subsection{Model comparison}

We rely on ROC curves to compare model performances, and in particular we compute for all distributions Youden's $J$ statistic to infer the threshold leading to optimal classification \citep[see][for example]{bewick2004statistics}: 

\begin{equation}
J = \frac{\text{TP}}{\text{TP}+\text{FN}}+\frac{\text{TN}}{\text{TN}+\text{FP}}-1
\end{equation}

where $TP$, $TN$, $FP$ and $FN$ account for the True/False Positives/Negatives.

Computing grapheme distances can be computationally expensive, especially with a large corpus of posts from social network \cite{Pacheco_2020}, thus we also evaluated the wall-time of each method with our implementation on a standard laptop. This enables to provide a relative order of magnitude for the time required by well-established implementations (namely, \textbf{polyleven} for Levenshtein and \textbf{difflib} for Ratclif-Obershelp) as well as our own Python implementation for the other algorithms.

Some methods have very close performance for our selection criterion, thus we computed 95\% confidence interval of the $J$ statistic by bootstrapping \cite{efron1986bootstrap}. 
Specifically, we generated 10,000 random bootstrap samples from the distributions of model performances per threshold, and compute the 95\% confidence interval of the empirical distribution true positive rates and false positive rates on these samples.

%%%%%%%%%%%%%%
%% DATASETS %%
%%%%%%%%%%%%%%
\subsection{Datasets} \label{datasets}

We study the general properties of our method using a newly developed, specifically crafted dataset to include translations, rewordings and copy-pastas. We also demonstrate how our method can be applied in the context of information manipulation based on a known campaign involving Venezuela in 2021, using a publicly available dataset by Twitter. %(\cite{url_by_twitter}).

\subsubsection{Synthetic dataset}

We design our synthetic dataset based on a corpus of short texts on which we then perform specific transformations matching the modus operandi of translations, rewordings and copy-pastas. We rely on seed datasets available on the Kaggle platform, chosen to encompass different registers while being representative of short messages posted on social media: the English disaster tweets dataset \cite{kaggletwitter}, a French books reviews \cite{kagglebook}, and headlines from the Huffington Post \cite{misra2022news}. 

To perform rewording and copy-pasta transformations, we employed ChatGPT\footnote{\url{
https://chat.openai.com}}, based on a generative language model \cite{brown2020language}. 
When instructing the model for rewording, we provided prompts that simply asked it to ``rephrase'' messages from the seed dataset, without specifying the exact nature of the rephrasing or including specific instructions. In the case of copy-pasta, our prompts explicitly guided the model to make specific alterations, such as ``Modify punctuation and add emojis in nine different ways'' or ``Add hashtags and insert up to one word in nine different ways''. Keeping prompts concise ensured better control over the model's output, enhancing the reliability of labeling.

We further used DeepL\footnote{\url{https://www.deepl.com/translator}} to generate translations into ten different languages : English, French, Spanish, German, Portuguese, Chinese, Italian, Russian, Turkish and Japanese. These languages are part of the set handled by USE \cite{yang2019multilingual}.

\begin{figure}[t]
\centering
\includegraphics[width=1\columnwidth]{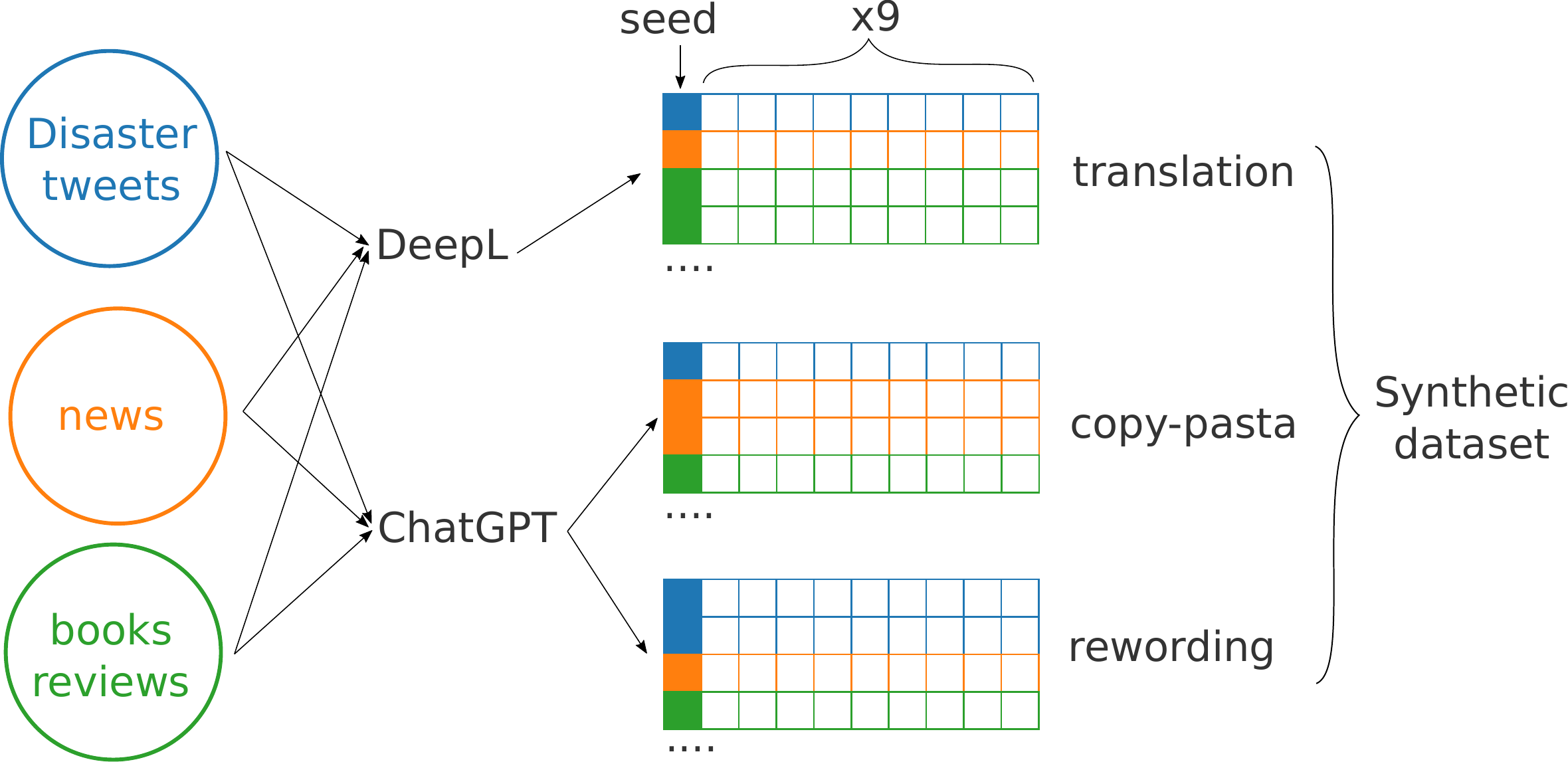}
\caption{Schema of synthetic dataset creation.}
\label{fig:dataset_schema}
\end{figure}

For each of the three seed datasets, we randomly selected 100 seed messages and applied all three transformations ten times, generating 3,000 pairs for each transformation (Fig. \ref{fig:dataset_schema}). Additionally, we included 1,000 pairs of messages randomly selected from each seed dataset to create non-matching message pairs with similar topics, forming our control set. In total, our dataset comprises 12,000 pairs of messages, covering all four conditions: 3,000 control, 3,000 copy-pasta, 3,000 rewording, and 3,000 translation.

\subsubsection{Twitter Transparency dataset}

To evaluate our approach, we use a multilingual dataset from Twitter Transparency Center denoted as \textit{Venezuela 2021}\footnote{\url{https://transparency.twitter.com/en/reports/moderation-research.html}}. This dataset refers to a moderation report publicly disclosed in December 2021.\footnote{\url{https://blog.twitter.com/en_us/topics/company/2021/disclosing-state-linked-information-operations-we-ve-removed}}  
According to the report, Twitter deleted 277 accounts that shared official Venezuelan government narratives using topic and hashtag amplification methods. Many of the accounts used proprietary Twitter clients to automate and amplify specific content. This dataset covers a broad period from 2009 to 2021. Similarly to all Twitter Transparency datasets, personal data are already anonymized (user ID, user name, user screen name, mentions, etc.). 

Here we focus on the users most recent activity from January to June 2021. As a result, our dataset encompasses a total of 249 accounts and 67,684 posts written in 28 languages, mainly in Spanish (86.2\%), but also marginally in English (1.5\%) and Portuguese (1.1\%). 75\% of the accounts were created after 2020. 

According to prior investigation of this dataset \cite{stanfordvenez}, the majority of hashtags and topics identified within the tweets are associated with political and commercial campaigns. Political-related tweets mainly revolve around two narratives : the liberation of Alex Saab (\texttt{\#AlexSaab}) - a businessman with affiliations to the Venezuelan presidential corruption system - and the support or denigration of several Mexican politicians (\texttt{\#KuriGobernador}, \texttt{\#RenunciaClaraLuz}, \texttt{\#YoConDavid}, etc.). 
Among those tweets, a substantial proportion openly displays political support for the gubernatorial candidate in the Mexican state of Querétaro, Mauricio Kuri González, elected as governor in October 2021. Some advocates for the resignation of Clara Luz Flores Carrales from her role as the Mayor of Escobedo in Mexico. 
The commercial tweets are promoting Mexican beverages, such as the Mexican beer brand cerveza Indio (\texttt{\#INDIOsustentable}) or Heineken's hard seltzer brand Pura Piraña (\texttt{\#SoyPuraPiraña}). A large campaign is about the diffusion of the TV show episode of The Walking Dead on STAR Channel, a privately-owned broadcaster (\texttt{\#TWDxSTARChannel}).

\subsubsection{Data preprocessing}

Generic data preprocessing entailed the removal of links and user mentions while retaining the hashtags. We also filtered out messages shorter than 30 letters (approximately 3-to-4 words), as we deemed them insufficiently informative to warrant duplication in the first place. In addition, for grapheme-distance processing, we also removed emojis, punctuation, spaces, and lowercased the text. This approach facilitated the process of identifying instances of copy-paste content, as the most straightforward ones were treated as identical strings from the outset.

In a broader context, our primary aim is to spot unusual patterns of repetition among users to uncover instances of coordination. Consequently, we calculate pairwise distances between individual account's timeline. This ensures that we do not mistakenly flag duplication within a single user, as it falls outside the scope of our research focus.

%\subsubsection{Code and Data Availability}

%All the code required to reproduce the analyses is made freely available on github\footnote{Note to reviewers: we will add the link to our github account here}. 
%The synthetic dataset is also made freely available on this repository. The Twitter transparency dataset is already available online. %\cite{twitter_download_link}.

%%%%%%%%%%%%%
%% RESULTS %%
%%%%%%%%%%%%%
\section{Results} \label{results}

\subsection{Synthetic Experiment}

To validate the effectiveness of our approach, we first conducted experiments using our specifically crafted synthetic dataset.

\begin{figure}
\centering
\includegraphics[width=0.9\columnwidth]{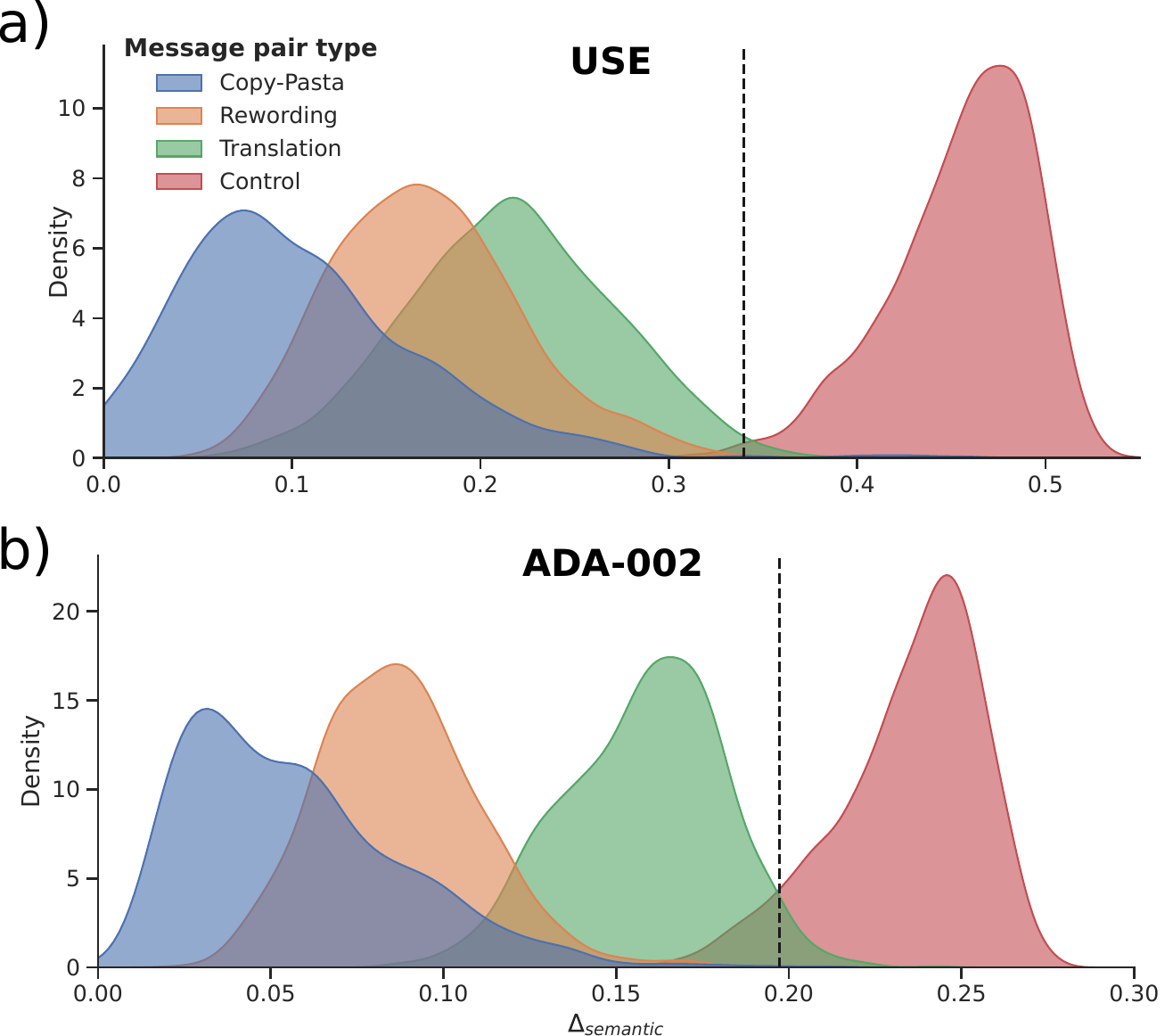}
\caption{
Density distribution of USE (a) and ADA-002 (b) distances for synthetic dataset message pairs. The optimal thresholds distinguishing between identical-meaning pairs (Copy-Pasta, Rewording and Translation) and the Control group are shown with dashed lines.}
\label{fig:dens_use}
\end{figure}

We evaluated the distribution of Universal Sentence Encoder (USE)-derived and ADA-002-derived distances between message pairs. The USE method effectively discriminated between messages possessing identical meanings (copy-pastas, rewordings and translations) and dissimilar meanings (controls), exhibiting distinct, non-overlapping distributions (Fig. \ref{fig:dens_use}a). Quantitatively, the USE separation was almost perfect (precision of 99.2\% ± 0.2\% and recall of 99.2\% ± 0.2\%, using the optimal threshold of 0.33). ADA-002 was less efficient in grouping together translations, which resulted in the distribution of translated message pairs being halfway between the controls and the rewordings (Fig. \ref{fig:dens_use}a). Quantitatively, this resulted in a precision of 95.2\% ± 0.6\% and recall of 95.2\% ± 0.6\% using the optimal threshold of 0.2.

\begin{figure}[!h]
\centering
\includegraphics[width=\columnwidth]{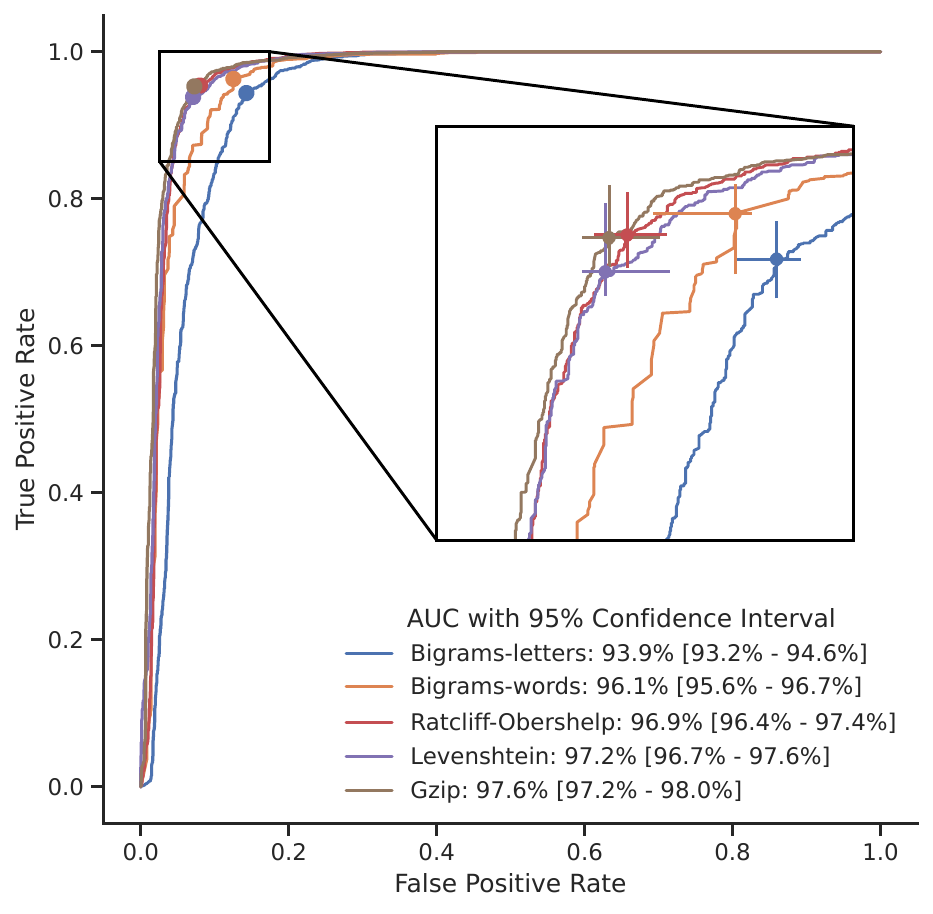}
\caption{
ROC curve benchmarking algorithm detection of Copy-Pasta \textit{vs.} Rewording. Markers correspond to the optimal solution according to the $J$ statistic. The 95\% confidence intervals computed from boostrapping are shown with error bars. Overall, the Gzip, Levenshtein, and Ratcliff-Obershelp trio exhibit similar performances on the synthetic dataset, and outperform the two bigram-based approaches.}.
\label{fig:roc_grapheme}
\end{figure}

The benchmarked algorithms efficiently detected copy-pasta on the synthetic data (Fig. \ref{fig:roc_grapheme}). Three metrics -- Gzip, Levenshtein, and Ratcliff-Obershelp -- delivered similarly optimal performances, their 95\% confidence intervals overlapping for Area Under the Curve (AUC), precision and recall (Table \ref{tab:grapheme_comparison}). Among these three methods, we selected the Levenshtein method as it is much quicker than the alternatives (Table \ref{tab:grapheme_comparison}). On our synthetic dataset, the best threshold for copy-pasta was found to be $\tau_p=0.31$. This means that given two strings sharing similar semantic content, they can be considered as Copy-Pasta provided that at most 31\% of the graphemes need to be deleted, swapped or inserted to transform one string into the another.

\begin{table}[!h]
\begin{tabular}{lrrrr}
\toprule
 & AUC & Precision & Recall & Timing \\
\midrule
$gz$ & 97.6 ± 0.4 & 94.1 ± 0.6 & 94.0 ± 0.6 & 33:34 \\
$lv$ & 97.2 ± 0.5 & 93.4 ± 0.6 & 93.3 ± 0.6 & 02:40\\ % 02:40
$ro$ & 96.9 ± 0.5 & 93.8 ± 0.7 & 93.6 ± 0.6 & 07:33:01 \\
% $lcs$ & 96.9 ± 0.5 & 93.6 ± 0.6 & 93.4 ± 0.6 & 62:09 \\
$bg/w$ & 96.1 ± 0.5 & 92.3 ± 0.7 & 91.8 ± 0.7 & 12:14\\
$bg/l$ & 93.9 ± 0.7 & 90.4 ± 0.7 & 90.0 ± 0.7 & 41:04\\
\bottomrule
\end{tabular}
\caption{
Goodness-of-fit comparison of various algorithms aimed at detecting Copy-Pasta vs Rewording. In this table, $gz$, $lv$, $ro$, $bg/w$ and $bg_l$ refers respectively to distances based on Gzip, Levenshtein, Ratcliff-Obershelp, bigram-words and bigram-letters distances (see Methods for details). The timing has been computed on a single core 2.4GHz CPU processing all pairwise distances on a dataset of 1000 messages (involving roughly 500,000 distances between messages). The leading methods (Gzip, Levenshtein, and Ratcliff-Obershelp) show overlapping confidence intervals, indicating similar performance. Levenshtein, the most time-efficient, completes this task in under three minutes.
}
\label{tab:grapheme_comparison}
\end{table}

\begin{figure}[!h]
\centering

\includegraphics[width=1.0\columnwidth]{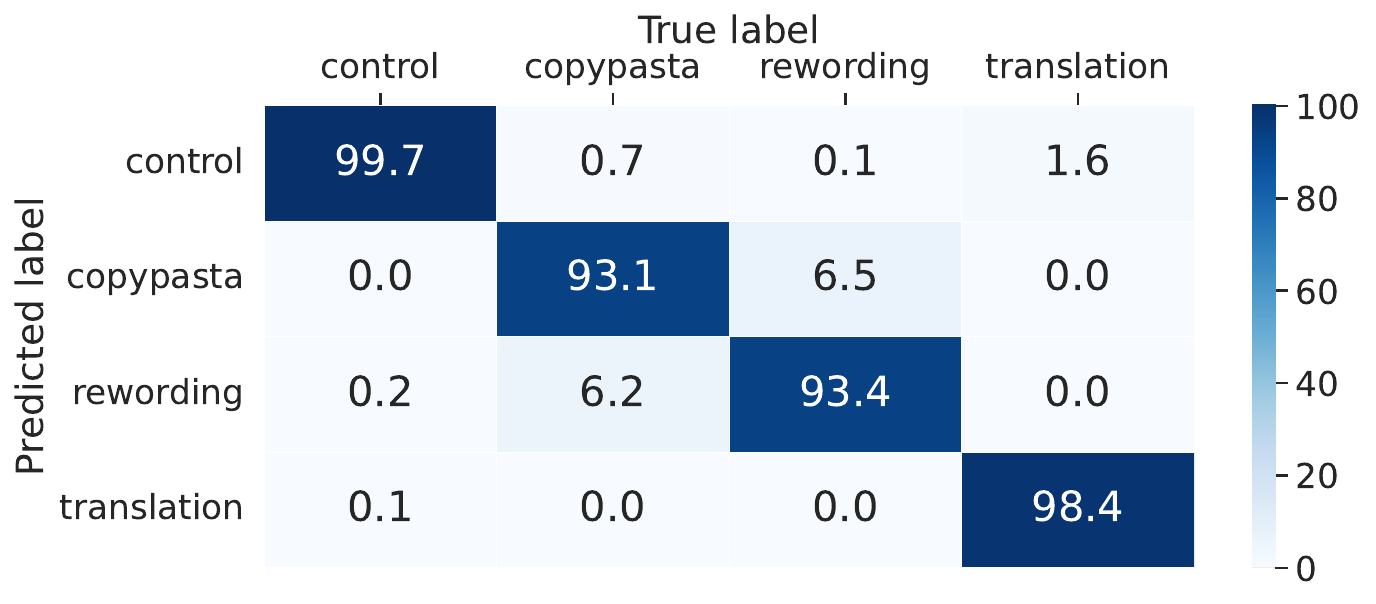}
\caption{Confusion matrix demonstrating algorithm performance on the synthetic experiment with semantic distance optimal threshold ($\tau_s=0.33$) and grapheme distance optimal threshold ($\tau_p=0.31$).}
\label{fig:conf_matrix_overall}
\end{figure}

\begin{figure*}[!t]
\centering

\includegraphics[width=1.8\columnwidth]{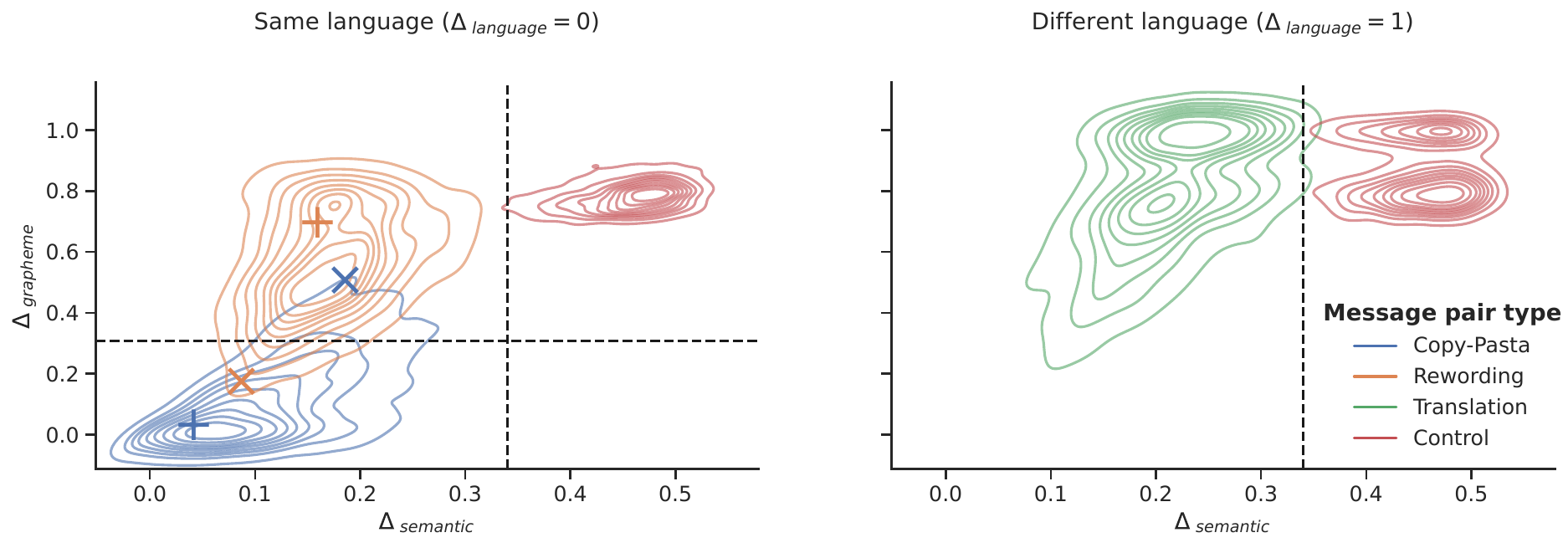}
\includegraphics[width=1.6\columnwidth]{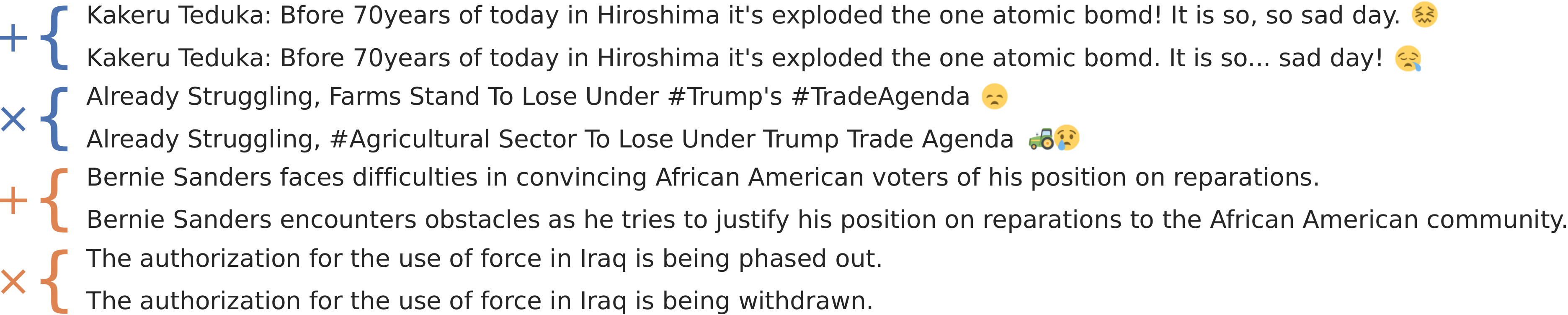}
\caption{
Distribution of the $3\Delta$-space distances computed for message pairs of the synthetic dataset. Plot densities represent each transformation, also including ``Control'' cases. 
The left plot shows the same-language condition, where the three possible categories are Copy-Pasta, Rewording and Control. The right plot shows the different-language condition and its two categories: Translation and Control. %Overall, the different distributions display high separability. 
A few representative examples are annotated in the plots and shown below, with ``plus'' glyphs denoting well-classified and ``cross'' glyphs denoting badly classified. The color code of the examples match the category, thus \textcolor{blue}{$\pmb +$} is a true positive for Copy-Pasta, and \textcolor{blue}{$\pmb\times$} is a false positive for Copy-Pasta.
As these examples illustrate, the relative overlap between the Copy-Pasta and Rewording can be attributed to ambiguous pairs whose ground truth labels could fall in either category.}
\label{fig:synthetic_2d}
\end{figure*}

Fig. \ref{fig:synthetic_2d} presents the overview $3\Delta$-space combining language, semantic, and grapheme distances. It demonstrates the methodology is well-suited to distinguish between Copy-Pasta, Rewording, Translation, and the Controls. Using the optimal thresholds, class accuracies were above 98\% for Control and Translation, and above 93\% for Copy-Pasta and Rewording (Table \ref{fig:conf_matrix_overall}). The slightly less good performance for the later classes highlight that they overlap to some extent in our dataset (Fig. \ref{fig:synthetic_2d}).

%\FloatBarrier

\subsection{Application on Real Data} % venezualian and mexican networks ?

Our evaluation of our methodology uses the \textit{Venezuela 2021} dataset. Following the data preprocessing described in Methods, we consider 21,071 tweets (31\%) and 239 users (96\%, Table \ref{tab:data_volume}). On this dataset, we applied the USE method due to its optimized semantic discrimination abilities, as well as the Levenshtein approach for its rapid and efficient results for grapheme distance. The threshold for Copy-Pasta/Rewording was set according to the value found in the synthetic experiment, i.e. $\tau_p = 0.31$. For minimization of false positives, we conservatively set the semantic threshold as $\tau_s = 0.2$ --- this aimed to prioritize precision over recall. We tested other threshold values, but kept this one for the analysis presented here, as it is sufficient to identify all the groups found in the prior work of \citet{stanfordvenez} while keeping graphs as light as possible. Applying these parameters, 177 users out of 239 (74\%) were found to engage in the abnormal duplication of 2,159 tweets (10\% of the dataset after post-preprocessing). These duplicate instances consisted of 9,381 copy-pastas, 1,047 rewordings, and 7,887 translations.

\begin{table}[h]
\begin{center}
\begin{tabular}{lcc}
\toprule
 & Users & Messages \\
\midrule
 Whole dataset          & 249    & 67684  \\
 Retweet deletion      & 244    & 39514  \\
 Length cleaning        & 239    & 21071  \\
 \textbf{Duplicates}       & \textbf{177}    & \textbf{2159}   \\
\bottomrule
\end{tabular}
\caption{
Overview of user and message numbers from the \textit{Venezuela 2021} dataset before and after preprocessing, along with those identified as duplicate by our methodology as copy-pasta, rewording, or translation.
}
\label{tab:data_volume}
\end{center}
\end{table}

% \subsubsection{Message clusters}

\begin{figure*}[t]
\centering
\includegraphics[width=1.85\columnwidth]{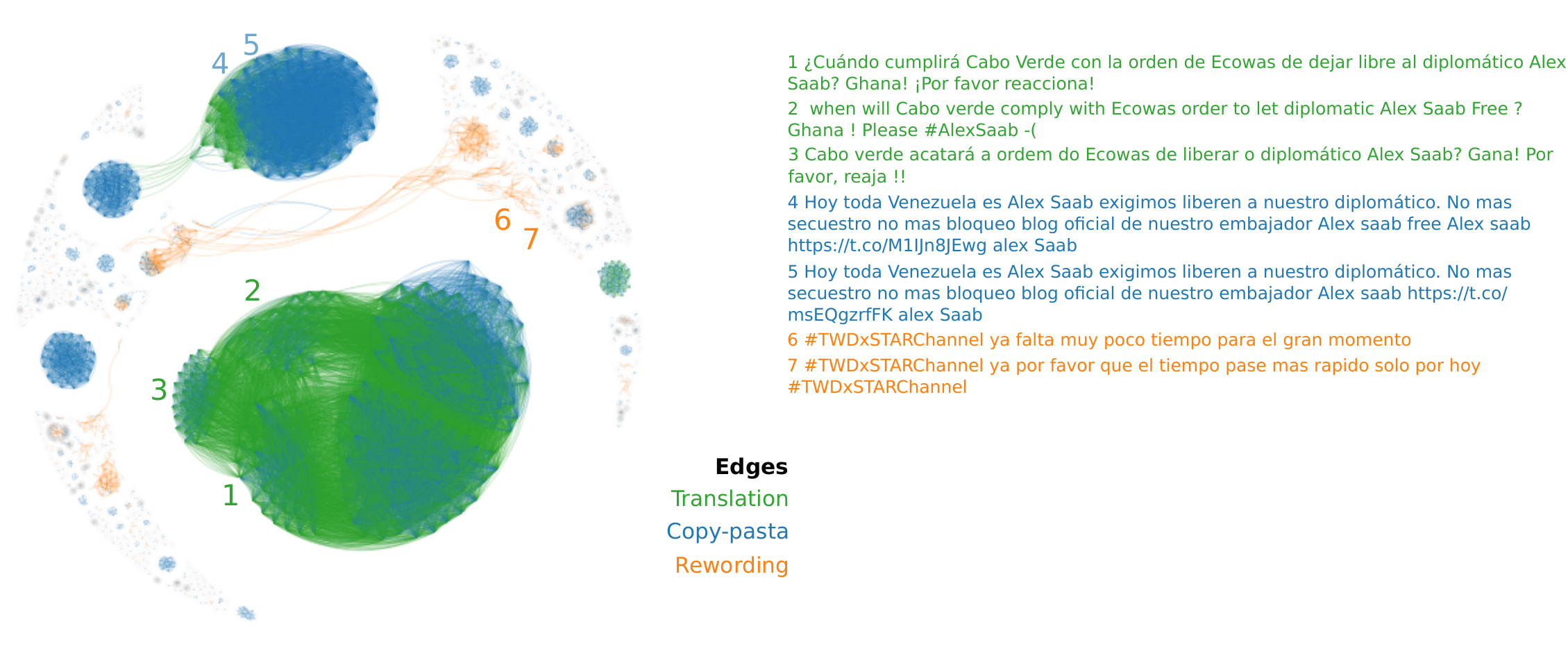} 
\caption{
Network representation of artificial duplicates within the \textit{Venezuela 2021} Twitter Transparency dataset. Messages are nodes, duplication methods are illustrated as colored edges. Some illustrative messages of Translation (green), Copy-Pasta (blue) and Rewording (orange) are numbered in the graph and reproduced on the right side.}
\label{fig:graphes4}
\end{figure*}

Consolidating individual transformations (Copy-Pasta, Rewording and Translation) at the network-level results in a graph with nodes representing tweets and edges denoting inauthentic duplication strategies (Fig. \ref{fig:graphes4}). The network's structure reveals distinct clusters that adopt specific modus operandi. The smaller clusters represent many isolated core messages duplicated minimally, often using a single modus operandi. The larger clusters account for the bulk of the dataset and represent a few narratives duplicated at large, and resort massively to Copy-Pasta.

\begin{figure*}[t]
\centering
\includegraphics[width=2\columnwidth]{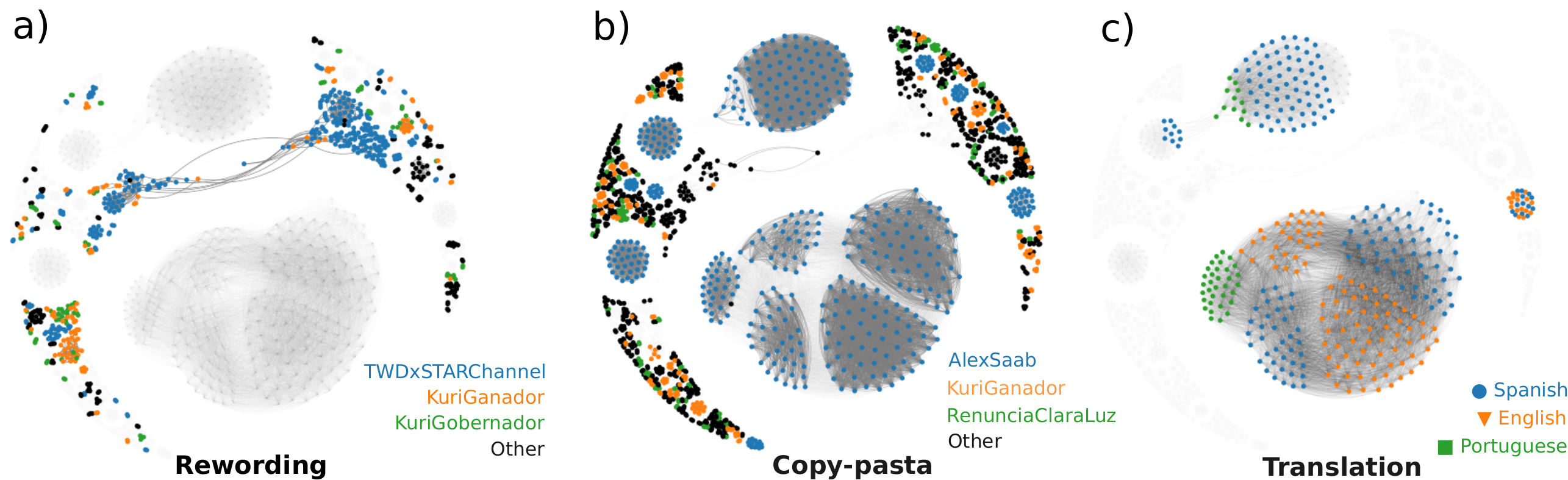} 
\caption{
Network visualizations broken down by duplication type. Nodes represent tweets, and edges represent Rewording (a), Copy-Pasta (b) and Translations (c). In each panel, the three most represented hashtags and languages are shown with a colorcode.
}
\label{fig:graphes3delta}
\end{figure*}

\begin{figure*}[!h]
  \includegraphics[width=2\columnwidth]{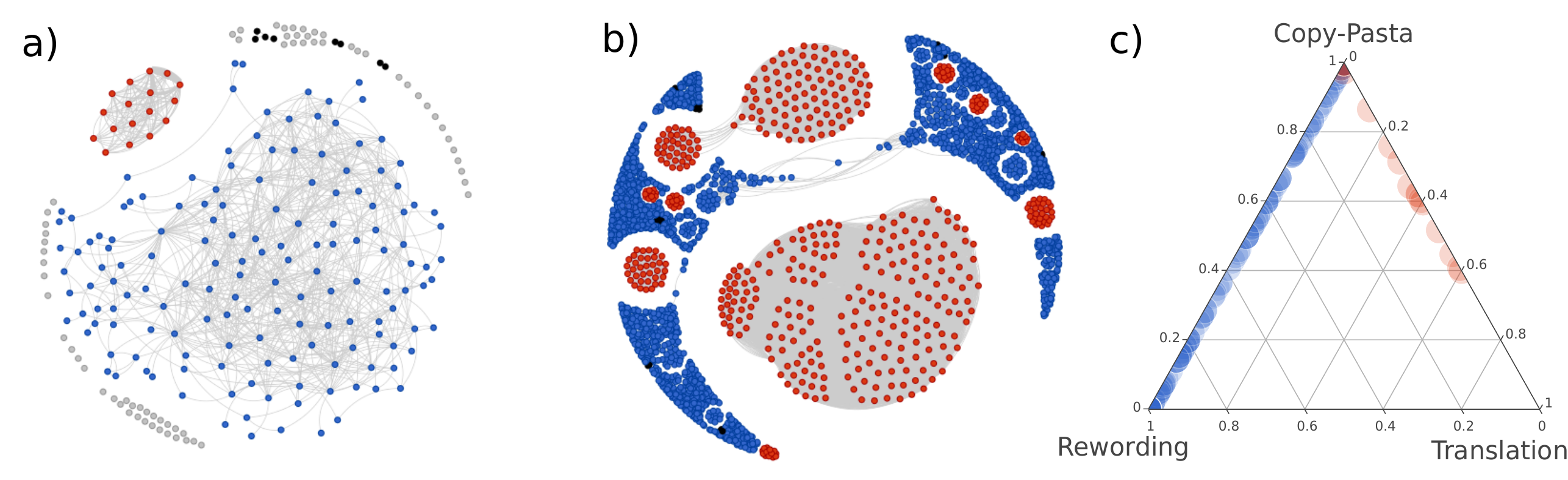}
\caption{
Account-level duplication analysis. Panel a) represents the network of accounts (nodes) linked if they share the same duplicate content (edges), regardless of the specific duplication method used. Panel b) represents the message-level network layout used in Figs. \ref{fig:graphes4} and \ref{fig:graphes3delta}, overlaying the account responsible for the tweet as a color code representing the clusters of panel a. Panel c) uses the same color code and maps accounts in a ternary plot according to the duplication methods they employ.
}
\label{fig:user_clusters}
\end{figure*}

Analyzing each modus operandi individually exposed their specific usage patterns within duplicated messages. For each strategy, we identified the top three most represented narratives or languages within duplicated tweets, which we utilized as labels for the nodes in Figs. \ref{fig:graphes3delta}a, \ref{fig:graphes3delta}b and \ref{fig:graphes3delta}c. Fig. \ref{fig:graphes3delta}a illustrates the rewording strategy, most prevalent in discussions around the TV show ``The Walking Dead,'' notably through the use of \texttt{\#TWDxSTARChannel} (48\% of reworded tweets). Secondary clusters primarily mentioned \texttt{\#KuriGanador} (17\%) and \texttt{\#KuriGobernador} (12\%), likely in support of Mauricio Kuri's gubernatorial candidacy. Copy-Pasta (Fig. \ref{fig:graphes3delta}b) was the most widespread duplication practice, and it predominantly circulated narratives involving Alex Saab, in an apparent effort to defend the convicted businessman. Small Copy-Pasta clusters also supported Mauricio Kuri with \texttt{\#KuriGobernador} (14\%) and revolved around the resignation of politician Clara Luz, with \texttt{\#RenunciaClaraLuz} (10\%). Translation (Fig. \ref{fig:graphes3delta}c) was most prevalent in the largest duplicated clusters, typically used to extend Spanish narratives to English and Portuguese. This method was restricted to messages discussing Alex Saab, implying a specific international focus for this campaign.

We subsequently analyzed duplication at the level of accounts. There are two main clusters of accounts operating autonomously from each other (Fig. \ref{fig:user_clusters}a). We identified the smaller cluster as an intricately interwoven network of 17 interconnected users, with each member duplicating content from every other within the group. The larger cluster was also slightly less interconnected, and comprised 152 accounts. Despite being smaller in size, the first cluster was the one spamming the largest clusters of duplicates, while the second account cluster was responsible of a myriad of tiny message clusters (Fig. \ref{fig:user_clusters}b). They match the two groups that were previously and independently identified in \citet{stanfordvenez}. Our analysis further reveals that these two groups exhibited a distinct combination of modus operandi, with the larger group employing both Copy-Pasta and Rewording, and the smaller favoring Copy-Pasta and Translation (Fig. \ref{fig:user_clusters}b). This differential signature may suggest that the two groups of accounts are operated by different handlers, or that they are using different operational processes or different software.

\begin{figure}
\centering
\includegraphics[width=\columnwidth]{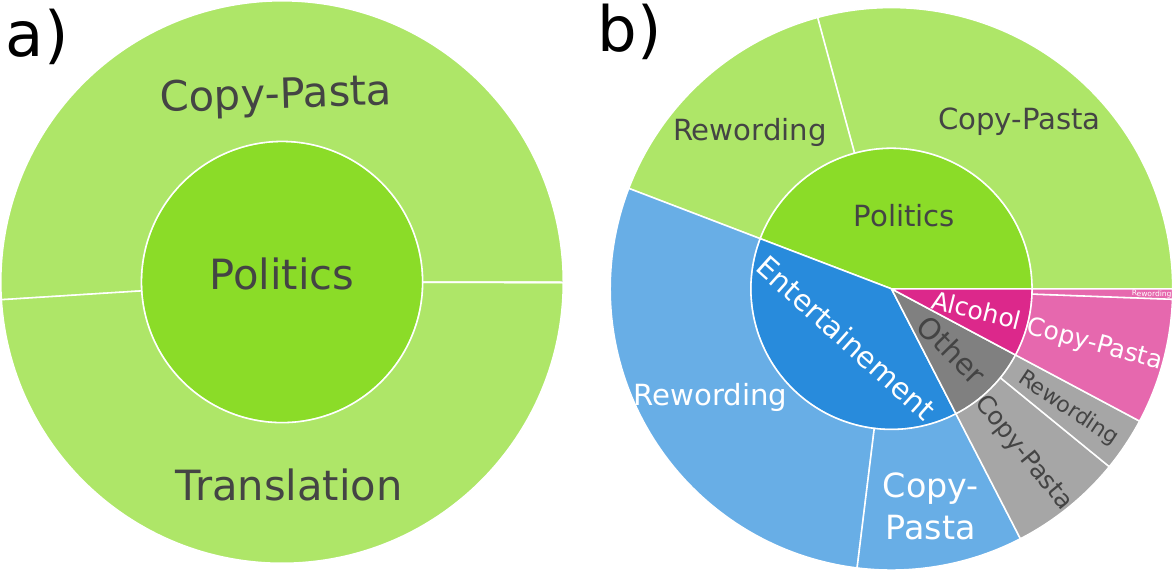} 
\caption{
Sunburst charts displaying duplication methods per topic discussed by the small account cluster (panel a, this cluster corresponds to the red nodes in Fig. \ref{fig:user_clusters}) and the large account cluster (panel b, corresponds to the blue nodes in Fig. \ref{fig:user_clusters}).}
\label{fig:sunburst_ven}
\end{figure}

For thematic analysis, we linked the topics discussed by the two user clusters with their respective duplication strategies (Fig. \ref{fig:sunburst_ven}). We categorized each message based on the main themes presented by their associated keywords in three general themes : politics\footnote{\#AlexSaab, Alex Saab, \#YoConDavid, \#KuriGanador, \#KuriGobernador, \#RenunciaClaraLuz, \#GobernadoraNoSeras, \#VamosBorrego, \#BrozoConSamuel, \#SamuelConBrozo, \#VoyConChristian}, entertainment\footnote{\#TWDxSTARChannel, \#LordVideoCentro} and alcohol\footnote{\#SoyPuraPiraña, \#INDIOsustentable}. This simple strategy was sufficient to label 93\% of the messages. The smaller cluster exclusively published politically charged content involving Alex Saab, employing an equal measure of Copy-Pasta and Translation (Fig. \ref{fig:sunburst_ven}). The larger cluster, conversely, disseminated more diverse content: commercials related to alcoholic beverage, political campaign messages supporting Mexican candidates, and discussions related to entertainments, particularly focused ``The Walking Dead.''

Lastly, we confirmed the validity of the $3\Delta$-space methodology by using traditional analysis, focusing on the two user clusters (Fig. \ref{fig:user_clusters}a) and based on account creation dates and specific posting behaviors. The smaller, more interconnected cluster displayed sequential activity with fixed posting times (22:00-04:00 UTC, or 18:00-00:00 in Venezuela), suggesting highly coordinated behavior. The massive repetition of narratives involving Alex Saab, their uniform user account characteristics (creation date between 4 and 9 May 2021, no description, location or followers, and an average of 10 followings) are indicative of coordinated inauthentic behaviour. The larger cluster's exhibited stereotyped posting behavior, successive waves spanning brief time frames of 1 to 2 hours. Beside, the amplified subjects have no real connection with each other and amplify systematically commercial or political interests, leading us to also validate our observations of artifical amplification by this network.

%%%%%%%%%%%%%%%%
%% DISCUSSION %%
%%%%%%%%%%%%%%%%
\section{Discussion}

Our method provides novel insights into the strategies employed by malicious networks to manipulate textual content. By clustering messages within the $3\Delta$-space, we have developed a precise method for identifying the nature of these alterations, categorizing them as Copy-pasta, Rewording, or Translations. Graph visualizations of clusters comprising altered messages or the users behind them, thus immediately reveal Coordinated Inauthentic Behavior. Consequently, our method initiates the investigative process directly by detecting coordinated inauthentic behavior, eliminating the need to rely on the identification of suspicious repeated hashtags or abnormal co-mentions in user-network analysis to characterize manipulation. 

To the best of our knowledge, this study represents the first comprehensive analysis of copy-pasta detection, and provides a dedicated dataset for further investigation of such a phenomenon. In our evaluation, we opted for the Levenshtein distance metric due to its strong interpretability and efficient performance. Surprisingly, the compression-based method Gzip also demonstrated good performance. The Ratcliff-Obershelp algorithm exhibited commendable results in our benchmark; however, it is non-commutative in its reference implementation and has notably slower execution compared to alternatives. The two original bigram-based methods we introduced underperformed significantly in our benchmark assessment, thus we cannot recommend their use. 
Note that our examination of Copy-Pasta was confined to languages that primarily use alphabets. Logographic writing systems, such as Chinese, feature shorter sentences with graphemes that convey more information. When dealing with such languages, it becomes essential to fine-tune the discrimination threshold for Levenshtein proximity by reducing it to ensure effective application.

Translation is an often overlooked aspect of information manipulation. The naive manual approach to detecting translations involves searching for translated strings to identify whether they were published in other languages. However, a single message can be translated into numerous variations, making this approach time-consuming and inefficient. 
Our approach, as exemplified in both our synthetic experiment and the \textit{Venezuela 2021} dataset, offers a reliable means of identifying content translated into different languages.
Our methodology holds particular value in the context of investigating translation across different social networks. This is especially relevant when it comes to tracking elements from preparation-platforms where content originates, typically created by smaller communities in their native languages, to exposition-platforms where the content is shared with the target audience in their language.

The proliferation of generative language models poses a substantial challenge in the battle against disinformation. These models have unlocked the potential for massively automated content creation, a trend already observed in practice.\footnote{see, e.g. \url{https://www.theverge.com/2023/4/25/23697218/ai-generated-spam-fake-user-reviews-as-an-ai-language-model}} However, determining whether individual messages are generated by AI remains a complex task, one that may even be insurmountable, as noted by \citet{sadasivan2023aigenerated} and \citet{yang2023anatomy}. Rather than focusing on individual messages, our approach operates at the cluster level, identifying messages that convey similar meanings but employ different phrasing. This approach shows great promise in detecting large-scale instances of AI generated text. In fact, to validate its effectiveness, we successfully applied it to a synthetic dataset primarily generated using ChatGPT. Beyond specific case studies such as the ones demonstrated here, our methodology holds broader applicability. For instance, it could be applied to the analysis of Twitter's 1\% random sample of messages flux, offering valuable insights into the changing prevalence of proliferation of AI-generated messages over time.

\bibliography{aaai22.bib}

\end{document}